t

\documentclass[aps,twocolumn,pre]{revtex4}
\usepackage[dvips]{graphics}
\usepackage{graphicx}
\usepackage{amsfonts}
\usepackage{amssymb}
\usepackage{amsmath}

\newcommand{\af}{\alpha}
\newcommand{\bn}{\bar n}

\newcommand{\la}{g}

\newcommand{\beql}[1]{\begin{equation}\label{#1}}
\newcommand{\eeq}{\end{equation}}
\newcommand{\eqn}[1]{(\ref{#1})}

\begin{document}
\title{
 Theory of Nonlinear Dispersive Waves and Selection of the Ground State}
\author{A. Soffer$^1$ and M.I. Weinstein$^2$}
\affiliation{
$^1$Mathematics Department, Rutgers University, New Brunswick, NJ 08903 USA \\ $^2$ Department of Applied Physics and Applied Mathematics, Columbia University,  New-York, NY  10027 USA}


\begin{abstract}
A theory of time dependent nonlinear dispersive equations of the Schr\"{o}dinger / Gross-Pitaevskii and Hartree type is developed.
The short, intermediate and large time behavior is found, by deriving  nonlinear Master
equations (NLME), governing the evolution of the mode powers, and 
 by a novel multi-time scale analysis of these equations. The scattering theory is developed and coherent resonance phenomena and associated lifetimes are derived. Applications include BEC large time dynamics and nonlinear optical systems. The theory reveals a nonlinear transition phenomenon, ``selection of the ground state'', and NLME predicts the decay of excited state, with half its energy transferred to the ground state and half to radiation modes. Our results predict the recent experimental observations of Mandelik {\it et. al.} \cite{MLS} in nonlinear optical waveguides.
\end{abstract}
\maketitle
Recent experimental \cite{EJMWC,DMAD,BSH,MLS} and theoretical
\cite{V-L,Sal,LSY,AFGST,Adh} breakthroughs in the field of
Bose-Einstein Condensates (BEC) and the intense interest in
nonlinear optical devices \cite{AHKW} necessitates the detailed
analysis of time-dependent nonlinear dispersive equations, such as 
the time-dependent Gross-Pitaevski (GP) and  nonlinear Schr\"odinger (NLS) equations in one or
more spatial dimensions. Also, new coherent phenomena in soft
condensed matter, such as macroscopic resonances \cite{BLS,W-S},
DNA-denaturation dynamics and macromolecule dynamics, have rekindled
the need to determine the large time behavior of time-dependent
Hartree - Fock (HF) type equations. In contrast to  linear scattering,
where time-independent methods are used to compute approximate $S$-matrix for general multichannel processes, nonlinear problems are not amenable to such theory, except in completely
integrable cases. Thus, mainly large scale numerical methods have been
applied so far. 

In this Letter we describe and apply a rigorous time-dependent theory and approach developed in \cite{S-W3} to provide detailed explicit  behavior for a class of nonlinear dispersive partial differential equations, which include NLS, GP and HF, on short, intermediate and infinite time scales. We show, as a consequence of our analysis, that in a multimode nonlinear dispersive systems which has multiple nonlinear bound states, that nonlinear resonant interactions lead to a crystallization of the coherent and localized part of the solution on the nonlinear ground state ({\it selection of the ground state}); the nonlinear excited states are metastable and decay at a rate given by a nonlinear analogue of Fermi's golden rule.

In very recent experimental work \cite{MLS} on  CW beams centered at wavelength, $\lambda=2\pi/k_0$,  propagating in multimode nonlinear AlGaAs waveguides, the output distribution of optical power is measured, and ground state selection and the partition law (\ref{partition}) have been demonstrated.
Our results predict these observations and can be used to interpret the earlier work on photorefractive waveguides \cite{Segev}.

 For definiteness we consider the nonlinear
Schr\"{o}dinger equation in three space dimensions:
\begin{equation}\label{NLS}
i \partial_t\phi = H_0 \phi +
\la F (|\phi|^2 ) \phi.
\end{equation}
 $F$ denotes the
nonlinearity, here taken to be local, $F(|\phi|^2)=|\phi|^2$, but more generally, of Hartree - Fock type type
$F(| \psi |^2) \equiv \int G(x,y) |\psi (y) |^2 dy$
with, for example, $G(x,y)=|x-y|^{- \af} e^{- r |x-y|},\ r\ge0$.  $\la$ denotes the coupling coefficient, related to the scattering length in BEC and, in optics, to $-n_2$, the nonlinear Kerr coefficient.
 $H_0=-\Delta+V(x)$, where $V(x)$  denotes a potential which decays to zero sufficiently rapidly at infinity, and
  for which $H_0$ has two bound
states (for simplicity); $H_0\psi_{j*}=E_{j*}\psi_{j*},\ j=0,1$. Such
equations appear in many applications in which coherent soliton-like structures and their internal modes \cite{KPCP} interact with dispersive waves, {\it e.g.} 
\cite{AFGST,AHKW,BLS,Ab,W-S,So}. The effective equations of an atom in a field or active
media is also of this form \cite{W-S,W-M}.
The requirement on $V(x)$ that $H_0$ have 
(at least) two bound states makes the model of general interest and the phenomena very rich.

Nonlinear bound state solutions bifurcate from  linear bound states  for weakly nonlinear perturbations \cite{RW} giving rise to {\it nonlinear ground state} family and {\it nonlinear excited state families} (for two bound state Hamiltonians). Specifically, 
for energies close to $E_{0\ast}$ and  $E_{1 \ast}$, the eigenvalues of $H_0$,
there are nonlinear
bound states: $e^{-iE_0 t} \psi_{E_0} , \quad \psi_{E_1} e^{-iE_1 t}$,
which solve (\ref{NLS}). $\psi_{E_0}$ and $\psi_{E_1}$ are complex-valued  (exponentially) localized solutions of $\left( H_0 + \la F(| \psi_{E_i} |^2) \right)
\psi_{E_i} = E_i \psi_{E_i}.
$
When $\la =0$, $E_0 = E_{0 \ast}$ and $E_1 = E_{1 \ast}$ are values for which there are nontrivial solutions. For $\la \neq 0$ there is a solution for all $E$ near $E_{i \ast}$.
$\psi_{E_j }$ can be thought of as a {\it pinned} soliton, ground ($j=0$) or excited state ($j=1$), within the support of $V(x)$. We assume $\omega_*\equiv 2E_{1*}-E_{0*}>0$, ensuring coupling of bound states to the continuum at second order; analogous results with slower decay rates hold more generally.

The main results we establish concern the time evolution of  solutions to (\ref{NLS}) for all data (with small energy).
The {\em large time behavior} is as follows: 
For any initial data, with small energy the solution $\psi (t)$ of
(\ref{NLS}) is asymptotically given by 
\begin{align}
\psi (t)\ &{\to}\ \ e^{-i \theta_0^\pm (t)} e^{- iE_j^\pm
t} \psi_{E_j^\pm}\ +\  e^{i \Delta t} \psi_\pm,\ \ {\rm as}\ t \to \pm \infty,
\nonumber
\end{align}
in $L^2$.
Generically, $j=0$, corresponding to the ground state, while there is a finite dimensional set of data for which $j=1$. Thus, for typical initial conditions we have {\it ground state selection}. We analyze the detailed  dynamics by reduction of the NLS to a finite dimensional system, 
which dominates the full dynamics.
 
%

The main second result concerns the transient, {\em finite} time,  behavior of the system, which is very rich:
There are {\em three} time scales, determined by the initial data.
In the first time interval, the dynamics is dominated by radiative
terms ({\it formative stage}); once the initial radiative part propagates
far enough from the support of $V(x)$, the soliton's position,
the second period ({\it embryonic stage}) begins. This is marked by a {\em monotonic} (exponentially fast) 
increase in the ground state amplitude, relative to the excited state amplitude. This stage proceeds as
long as the excited state part is larger than some fraction of the
ground state part. Once the ground state reaches a size comparable
with some (fixed) fraction of the excited state, the third and
final stage begins ({\it ground state selection}); the ground state amplitude increases
monotonically and the excited state {\em decreases monotonically},
but at a polynomial rate as $t\to\infty$. (This describes the
generic case; the asymptotic state is given by an excited state
\cite{T-Y} whenever the first stage persists to time equal infinity.)
  A different approach, based on linearization around
the excited state for intermediate times and around the ground
state for large times, gives similar results (Soffer-Weinstein, unpublished notes, 1996).

 This phenomena is quantified by our derivation of the
{\em Nonlinear Master equations} (NLME)
   that govern the dynamics of the coupled ground and excited state modes (and the radiation modes $\ldots$).
If we denote by $P_0 (t)$ and $P_1 (t)$, respectively,  the (up to near identity transformations) squared projection of the system's state onto the ground state and excited states, at time $t$, we have
\begin{align}
\partial_t P_0 & =  2\Gamma P_0 P_1^2 +
\rho_{0}(t)
P_0 P_1^{M_0} + {\cal O}( t^{-5/2}  ) \nonumber\\
 \partial_t P_1 &
= - 4 \Gamma P_0 P_1^2 + \rho_{1} (t) \sqrt{P_0}
P_1^{M_1} + {\cal O}( t^{-5/2 } ),
\label{nlme}\\
& \Gamma = \pi \la^2\ \left|\ {\cal F}\left[dF[\psi_{0*}\psi_{1*}]\psi_{1*}\right](\omega_*)\ \right|^2\nonumber\\
& dF[\psi_{0*}\psi_{1*}]\psi_{1*} =\int G(x-y)\psi_{0*}(y)\psi_{1*}(y) dy\ \psi_{1*}(x).
\label{Gamma}
\end{align}
$\rho_{M_0} (t)$ and $\rho_{M_1} (t)$ are small oscillatory functions of time, $M_0,M_1\ge3$ and ${\cal F}[q](\omega)$ denotes the projection of $q$ onto the generalized eigenfunction of $H_0$ at frequency $\omega$. The
crucial number $\Gamma$ is (generically) positive and given
explicitly in terms of known eigenstates of $H_0$.
 $\Gamma \neq 0$ is the nonlinear Fermi Golden rule for such
systems and it gives the rate of {\em decoherence} and relaxation
\cite{S-W2,So,W-M,Wei,W}. The behavior of these Master Equations
(\ref{nlme}) reflects the three time scales mentioned above, on
which the behavior is very different. For large enough time, in
the third time domain, the last two terms in (\ref{nlme}) can be
ignored. In this latter regime, it is easy to see that $P_1(t)\to0$ as $t\to\infty$ and $\partial_t\left(2 P_0 + P_1 \right) =0$. Therefore, 
$2P_0 (\infty ) = 2 P_0 (t_0 ) + P_1 (t_0)$. It follows that
\begin{equation}
 P_0 (\infty) = P_0 (t_0) + \frac{1}{2}P_1 (t_0) ;
\label{partition}
\end{equation}
half of the excited state energy flows to
the ground state (the other half goes to radiation), a kind of {\em energy equipartition}. (The factor 2 also appears in the linear analysis,
and is interpreted as the ratio of relaxation to decoherence time
\cite{W-M,Wei}.) 
Furthermore, it follows that for any initial state $(P_0 (t_0),P_1(t_0))$ with $P_0(t_0)\neq 0$ the system converges to $(P_0(\infty),0)$ with a rate $\sim\ (1+ 4\Gamma P_0 (\infty ) P_1(t_0) t)^{-1}$  (selection of the ground state).  
Combining the analysis of the NLME (\ref{nlme}) for
all time scales with the previous statements gives a complete
description of the solution for all time scales and, in particular,
the asymptotic behavior, relaxation and decoherence rates, the
asymptotic profile and energy of the soliton/ground state.

 We sketch our method, for the special case of (\ref{NLS}) with cubic nonlinearity, $F(|\phi |^2)\phi=|\phi|^2\phi$.
Our approach makes use of ideas from  \cite{S-Wgafa, S-W1,
 S-W2,BPCucc,S-W3}.
 We begin with the Ansatz $ \phi (t) \equiv e^{-i \theta (t)} [ \psi_0
(t) + \psi_1 (t) + \phi_2 (t) ]$
where $\psi_0 (t) \equiv \psi_{E_0 (t)}$ is a solution of the ground state eigenvalue equation
with energy $E_0 (t)$, at time $t$. $E_0 (t)$ will be determined
later by orthogonality conditions \cite{S-W1,S-W2,S-W3}. Similarly
$\psi_1 (t)$ is an excited state eigenvector with eigenvalue $E_1
(t)$.
$\theta (t) \equiv \theta_0 (t) + \tilde{\theta} (t),\ \
\theta_0 (t) = \int_0^t E_0 (s) ds$.
$\tilde{\theta} (t)$ will be chosen
appropriately;
it includes (logarithmic) divergent phase.
Substitution of the above Ansatz for $\phi$  into \eqn{NLS}, and complexifying
the equations $(\phi_2 \to (\phi_2, \bar\phi_2 ) \equiv \Phi_2 (t)$,
$(\psi_j \to (\psi_j, \bar\psi_j ) \equiv \Psi_j(t)$
etc.) we derive
\begin{align}\label{eq43}
&i \partial_t \Phi_2 (t) = {\mathcal
H}_0 (t) \Phi_2 (t) - i \partial_t \Psi_0\nonumber\\
& -\
\left[ (( E_0 - E_1 ) + \partial_t \tilde{\theta} ) \sigma_3 + i \partial_t \right]\ \Psi_1 +\vec{F}_{NL},
\end{align}
where $\vec{F}_{NL}$ is nonlinear in $\Phi_2$, $\Psi_0$,
$\Psi_1$, $\tilde{\theta}$ and ${\cal H}_0(t)$ is given by the matrix operator
\begin{equation}
\sigma_3 \left(
\begin{array}{cc}
H-E_0 (t) + 2 \la |\psi_0 (t) |^2 & \la \psi_0^2 (t) \\ [+.2in]
\la \bar{\psi}_0^2 (t) & H- E_0 (t) + 2 \la | \psi_0 (t) |^2
\end{array}
\right),\nonumber
\end{equation}
where $\sigma_3$ is the Pauli matrix ${\rm diag}(1,-1)$.
 We consider the spectrum of ${\mathcal H}_0 (t)$ for fixed $t$ and $| \psi_0 | \equiv |\alpha_0 |$ small:\ (a)\
The continuous spectrum extends from $-\mu$ to $-\infty$, and $\mu$ to $\infty$ where $\mu \equiv E_1 - E_0 + O (| \alpha_0 |^2 )$.
The discrete spectrum is $\{0, - \mu, \mu \}$, with $0 < |\mu | < |E_0 |$ by assumption.\ (b)
Zero is a generalized eigenvalue of ${\mathcal H}_0$, with  generalized eigenspace.
spanned by
$\left\{ \sigma_3 \Psi_0,\partial_{E_0}\Psi_0\right\}$.

The discrete spectral subspace has dimension four. Therefore, $\Phi_2$ which lies in the continuous spectral part of ${\mathcal H}_0 (t)$, is constrained by four orthogonality conditions. Furthermore, $\partial_t \tilde{\theta}$ 
is chosen to remove divergent logarithmic phase
 contributions.  
In the weakly nonlinear (perturbative) regime, bound states have expansions $ \psi_{E_j} 
= \alpha_j (\psi_{j \ast} (x) + \la |\alpha_j |^2 \psi_j^{(1)} (x) + {\cal O}(\la^2 |\alpha_j |^4 ))$ and
$E_g = E_{j \ast} + {\cal O} (| \alpha_j |^2 )$.
The system for $\Phi_2$ and $\vec\alpha=(\vec\alpha_0,\vec\alpha_1)$ can be written in the form:
$i \partial_t \vec{\alpha}\ =\ {\mathcal A} (t) \vec{\alpha} + F_\alpha ,\ \
i \partial_t \Phi_2\ =\ {\mathcal H} (t) \Phi_2 + F_\Phi ~.$

 To study the energy exchange between $\alpha_0$ and $\alpha_1$,
 it is important to express $\alpha_1(t)$ as a slow amplitude modulation of a rapidly varying phase. With this goal in mind, we study
 the equation for $\alpha_1$, expressible as
$i \partial_t \vec{\alpha}_1 = \vec{A} (t) \vec{\alpha}_1 + \vec{F}.$
Freezing $t$ at some arbitrary large time $T$ in $\vec{A} (t)$, we solve $i \partial_t \vec{\alpha}_1 = \vec{A} (T) \alpha_1$ which gives a periodic Floquet solution matrix $\bar{X} (t)$.
We use that to eliminate this (fast) oscillation in $\vec{\alpha}_1 \equiv X (t) \vec{\beta}_1$.
The difference
$\vec{A} (t) - \vec{A}(T)$ and other similar differences are higher order corrections, uniformly in $t$, as $t$, $T \to \infty$,
and therefore can be neglected.

To proceed further we decompose $\Phi_2$ into its continuous spectral (dispersive) part, $\eta$, relative to ${\mathcal H}_0(T)$, and its components along the discrete modes. The latter are higher order and controllable. Thus  NLS, at low energy is equivalent to a system of the form:
\begin{eqnarray*}\label{coupled}
i \partial_t \eta & = &
{\mathcal H}_0 (T) \eta + {\mathcal F}_\eta (t; \alpha_0, \beta_1, \eta ) \\
i \partial_t \beta_1 & = & 2 \la \langle \psi_{0 \ast} , \psi_{1 \ast}^3 \rangle | \beta_1 |^2
\alpha_0 e^{i (E_{1*}-E_{0*}) t}\\
&& + 2 \la
\langle \psi_{0 \ast} \psi_{1 \ast}^2, \pi_1 \Phi_2 \rangle
\bar{\beta}_1 \alpha_0 e^{2i (E_{1*}-E_{0*}) t}\ +\ {\cal R}_0 \\
i \partial_t \alpha_0 & = & \la \langle \psi_{0 \ast}^2 , \psi_{1 \ast}^2 \rangle e^{- 2i (E_{1*}-E_{0*}) t} \beta_1^2 \bar{\alpha_0} \\
&+&\ \la \langle \psi_{0 \ast} \psi_{1 \ast}^2 , \Phi_2 \rangle \beta_1^2 e^{- 2 i (E_{1*}-E_{0*}) t}\ +\ {\cal R}_1,
\end{eqnarray*}
where ${\cal R}_j$ denotes corrections of a similar form and higher order.

The above system can be viewed as an infinite dimensional  Hamiltonian system consisting of two subsystems: a finite dimensional subsystem governing ``oscillators'',
 $(\alpha_0,\beta_1)$, and an infinite dimensional subsystem governing the field, $\eta$. Although this system has time-rapidly varying coefficients and no evident direction of energy flow, we claim we have now made explicit the key aspects, which give rise to resonant energy exchange. This is made explicit via a detailed analysis of non-resonant and resonant terms
 {\it Non-resonant} oscillatory dependence contributions can be transformed by successive near-identity changes of variables, $(\alpha_0,\beta_1)\mapsto (\tilde\alpha_0,\tilde\beta_1)$, to higher order order
 in the energy of the data (assumed small) and perturbatively controlled. {\it Resonant} oscillatory terms cannot be transformed to higher order and contribute to the  finite dimensional reduction.

To arrive at the reduction, we  solve the  $\eta$ equation, making explicit all terms through second order in $\la$, using the Green's function $G(t,t') = e^{-i {\mathcal H}_0 (T) (t-t' )}$.
We  focus on the key terms coming from the sources in ${\mathcal F}_\eta$ or the type $\alpha_0^i \alpha_1^j$, $0 \le i,j \le 2$ and
having oscillatory phases $e^{i m_{ij} t}$.
Their contribution to $\eta$ is of the form
\begin{equation}
\sim \int_0^t e^{-i {\mathcal H}_0 (T) (t-t') } |\chi \rangle
e^{im_{ij} (t')} \alpha_0^i (t') \alpha_1^j (t' ) dt'\
\label{resint}
\end{equation}
where $\alpha_0 , \alpha_1$ is a component of either $\vec{\alpha}_0$ or $\vec{\alpha}_1$, where $| \chi \rangle$ is an  (exponentially localized) function of position, expressible in terms of $\psi_{0*}$ and  $\psi_{1*}$.
We insert this solution into the $\alpha_0$, $\alpha_1$ equations, in place of $\Phi_2$.
We obtain integro-differential equations for $\alpha_0$,
$\alpha_1, (\beta_1)$. Terms of the form (\ref{resint}) are solutions
to a forced linear system and among the forcing terms made explicit in  (\ref{coupled}) are oscillatory terms with the frequency $\omega_*$, which are resonant with the continuous spectrum.
 Internal dissipation resulting in nonlinear resonant energy transfer from the excited state to the ground state and to dispersive radiation  is derived from  these resonant terms; see also the derivation of 
internal dissipation in both linear and nonlinear
resonance theories, recently developed by us \cite{S-Wgafa, S-W2,So,W}. This dissipation coefficient is
 $\Gamma$, the rate of decoherence and relaxation.
The above described scheme gives
$i \partial_t \tilde{\alpha}_0  =  ( - \Lambda + i \Gamma ) |
\tilde{\beta}_1 |^4 \tilde{\alpha}_0 + \tilde{{\mathcal R}}_0 (t),\ \
i \partial_t \tilde{\beta}_1 =  2 ( \Lambda- i \Gamma ) | \tilde{\alpha}_0 |^2
| \tilde{\beta}_1|^2 \beta_1 + \tilde{{\mathcal R}}_1 ( t).$

Introducing the squared projections of the system's state onto the ground state and excited states, $P_0 \equiv | \tilde{\alpha}_0 |^2$,
$P_1\equiv | \tilde{\beta}_1 |^2 $ we obtain NLME, (\ref{nlme}).
The system (\ref{nlme}) is analyzed in terms of {\it renormalized} powers, $Q_0$ and $Q_1$, for which it is shown that  there exist transition times $t_0$ and $t_1$, such that: $Q_0(t)$ decays rapidly on $[0,t_0]$, $Q_0(t)/Q_1(t)$ grows rapidly on $[t_0,t_1]$ and 
 the finally on $[t_1,\infty)$ the following system governs: 
 $ \partial_tQ_0 = 2\Gamma Q_0 Q_1^2,\ 
\partial_tQ_1 = - 4\Gamma Q_0 Q_1^2$. This gives $Q_0 \uparrow Q_0 (\infty )$ and $Q_1 \downarrow 0$ at rates discussed above.

The decay of the nonlinear excited state can also be understood as a linear instability. Let ${\cal H}_1$ denote the linearization about the nonlinear {\it excited state}, the operator  in (\ref{eq43}) with $\psi_0$ replaced by the excited state $\psi_1$. Since $\omega_*>0$,
in the  limit  of vanishing nonlinear terms, ${\cal H}_1$ has an embedded discrete eigenvalue in the continuous spectrum. Perturbation theory of embedded eigenvalues (see, for example, the time-dependent approach in  \cite{S-Wgafa} ), can be applied to show that this embedded eigenvalue perturbs to a linear exponential instability
with exponential rate $\sim\Gamma$ associated with the linear propagator $\exp(-i{\cal H}_1\tau)$. 

 In the context of nonlinear optical waveguide experiments, {\it e.g.}~\cite{MLS}, our analysis gives the distance, $L_{\rm transf}$,  over which nonlinear resonant energy transfer occurs as:  $L_{\rm transf}^{-1}=4\Gamma P_0(\infty) P_1(0)$. 
Here $g=-n_2/\bn_0$, $n_2$\ $[m^2/W]$ the Kerr nonlinear coefficient, $P_j\ [V^2/m^2]$ the square of the electric field projection onto  state $j$. $\Gamma$ is obtained from (\ref{Gamma}), where the states $\psi_{j*}$ are those derived from the Hamiltonian, $H_0=-2k_0^{-2}\Delta_{x}+V(x)$, with potential $V(x)=(1/2)(1-(n_0(x)/\bn_0)^2)$,
 with $x$, the transverse coordinate.
 $n_0(x)$ and $\bn_0$ are, respectively, the linear refractive index and its value at infinity. This implies 
$L_{\rm transf}^{-1}\sim \left(4\pi k_0 \left(\frac{2n_2}{\varepsilon c\overline{n_0}^2} \right)^2\cdot 
 \left(\frac{Pwr}{{\rm A}_{eff}}\right)^2\cdot \left| {\cal F}[\psi_{0*}\psi_{1*}^2](\omega_*)\right|^2\right)$.
 For parameter values at wavelengths $\sim 1.5\mu m$ and peak power levels at $10^3\ W $, $L_{\rm transf}$ of device size dimensions ($100\ \mu m$ to $ 1\ mm$) can be attained. The  factor ${\cal F}[\cdot](\omega_*)$, which can be approximated by WKB, depends on the density of states 
 of $H_0$ near $\omega_*$,  and can be tuned by varying the waveguide material and geometric parameters.  
 A second application to optical devices 
is the use of periodic photonic microstructures with appropriately
designed defects to trap coherent light pulses.   For example,  it has
been shown theoretically that (Kerr) nonlinear and periodic
structures support gap solitons \cite{CJAW} and trap pulses
 \cite{YSWF} traveling at any speed
less than the speed of light. Experiments have demonstrated
soliton propagation at about $50\%c$ \cite{EggSlushBroderick} and
theoretical studies \cite{GSWGHW} show that gap solitons can be
trapped with appropriately designed defects. This has potential applications to optical buffering, high-density storage and optical gates. The light stopping mechanism, is based on the transfer of
energy from incoming solitons to the pinned nonlinear modes of the
defect. The ground state selection phenomenon and the partition of excited state energy into pinned nonlinear ground state and radiation modes (\ref{partition}), described by NLME, quantify the rate of transfer and efficiency of trapping.
For systems of BEC droplets in a potential well, the relaxation time to the ground state and the decoherence time due to the
presence external perturbations, {\it e.g.} other droplets, are measurable  and of fundamental importance related to $\Gamma$. For example, they are relevant to the construction and feasibility of quantum gates and memories.

Our analysis is applicable in Hamiltonian systems which can be  decomposed into a lower dimensional dynamical system (oscillators) coupled to an infinite dimensional dynamical system (wave-field),  acting as a dispersive ``heat bath'', {\it e.g.} how an open quantum system is effected by its environment. $\Gamma$ is related  to the rate of energy transfer or decoherence.  
In general, when a mean-field type
approximation to the interaction with the environment describes the system, the above analysis can be used, {\it e.g.} to study wave function collapse in various systems.

To summarize, we have derived the behavior on all short, intermediate and infinite time
scales of NLS type equations, in which general solutions involve an interaction among two families of nonlinear bound states (soliton-type) and dispersive radiation. We have derived Nonlinear Master equations, which govern the dynamics. The phenomena of
ground state selection was demonstrated, the rate of decay/relaxation  was computed, and the energy distribution of the asymptotic state of the system was derived. These phenomena were recently observed experimentally in [11]. Finally, we have shown the applicability of our theoretical results to nonlinear optical devices and to BEC dynamics and decoherence phenomena.  

\noindent  We thank D. Mandelik, Y. Lahini and Y. Silberberg, who  shared their manuscript  on their experimental work \cite{MLS} before publication and C-W Wong for very informative discussions. A.S. and M.I.W. were supported, in part, by grants from the US National Science Foundation.

\end{document}